\documentclass[a4paper,twoside,12pt,dvips]{article}
\usepackage{a4p}
\usepackage{cite,mcite}
\usepackage{graphicx}
\usepackage{physics}
\usepackage{l3_title}
\usepackage{ifthen}
\usepackage{epsfig,rotating}
\usepackage{amssymb}
\usepackage{pennames}

\journalname{Phys.~Lett.~B}
\date{June 27, 2000}
\preprint{2000-107}
\newlength{\capindent}
\newlength{\capwidth}
\newlength{\figwidth}
\newcommand{\icaption}[2][!*!,!]{\hspace*{\capindent}%
  \begin{minipage}{\capwidth}
    \ifthenelse{\equal{#1}{!*!,!}}%
      {\caption{#2}}%
      {\caption[#1]{#2}}
  \end{minipage}}
\newcommand{\pho}{\phantom{0}}

\newcommand{\PZ}{\ensuremath{\mathrm{Z}}}%      Z not in pennames
\newcommand{\Pgn}{\ensuremath{\nu}}%            generic neutrino not in pennames
\renewcommand{\Pqb}{\ensuremath{\mathrm{b}}}%   b replace q_b in pennames
\renewcommand{\Paqb}{\ensuremath{\bar\mathrm{b}}}%   b replace q_b in pennames
\def\dd#1{\ensuremath{\,{\mathrm{d}#1}}}%

\newcommand{\PARJ}[1]{{\scshape parj(#1)}}

\newcommand{\KORALW}{{\scshape koralw}}
\newcommand{\PYTHIA}{{\scshape pythia}}

\newcommand{\GEANT}{{\scshape geant}{\small 3}}
\newcommand{\GHEISHA}{{\scshape gheisha}}

\newcommand{\WW}{{\PW\PW}}
\newcommand{\interW}{inter-\kern -0.1em \PW}
\newcommand{\intraW}{intra-\kern -0.1em \PW}

\def\data{\ifmath{{\mathrm{data}}}}
\def\mix{\ifmath{{\mathrm{mix}}}}
\def\noBE{\ifmath{{\mathrm{noBE}}}}
\newcommand{\MC}{\ensuremath{{\mathrm{MC}}}}
\def\gen{\ifmath{{\mathrm{gen}}}}
\def\det{\ifmath{{\mathrm{det}}}}

\newcommand{\BE}{Bose-\kern -0.1em Einstein}

\begin{document}

\bibliographystyle{l3stylem}
\begin{titlepage}
%
%%%%%%%%%%%%%%%%%%%%%%%%%%%%%%%%%%%%%%%%%%%%%%%%%%%%%%%%%%%%%%%%
\title{Measurement of \\ 
       \BE\ Correlations      \\
%       Bose-Einstein Correlations      \\
       in {\boldmath$\Pep\Pem\rightarrow\PWp\PWm$ at $\sqrt{s}\simeq189\GeV$}
                 \\[-1cm]}
\author{The L3 Collaboration}

%
% The abstract
%
\begin{abstract}    \noindent
We investigate \BE\ correlations (BEC) in \PW-pair production
at $\sqrt{s}\simeq189 \GeV$  using the L3 detector at LEP.
We observe BEC between particles from a single \PW\ decay in good agreement
with those from a light-quark \PZ\ decay sample.
We investigate their possible
existence between particles coming from different \PW's.
No evidence for such  \interW\ BEC is found.
\end{abstract}
%
% Adds "To be submitted to ..." or "Submitted to ...", if relevant
%
\submitted
\end{titlepage}

%\newpage
%{\pagestyle{empty}
% \cleardoublepage
%}
\setcounter{page}{1}
%
%
%%%%%%%%%%%%%%%%%%%%%%%%%%%%%%%%%%%%%%%%%%%%%%%%%%%%%%%%%%%%%%%%%%%%%%%%%%%%%%%
% Introduction
%%%%%%%%%%%%%%%%%%%%%%%%%%%%%%%%%%%%%%%%%%%%%%%%%%%%%%%%%%%%%%%%%%%%%%%%%%%%%%%
%

%\include{kernel}
\section*{Introduction}

\BE\ (BE) interference is observed 
as an enhanced production of identical bosons, \eg, charged pions, at small four-momentum difference
in elementary particle and nuclear collisions \cite{Gold59,Boal90},
and, in particular,
in hadronic \PZ\ decay~\cite{opal-40,l3-174}. 
Such an interference should also be present in hadronic \PW\ decay  (\intraW\ BE interference).
Furthermore, since in fully hadronic \WW\ events
($\Pep\Pem\rightarrow\PWp\PWm\rightarrow \Pq\Paq\Pq\Paq$),
the \PW\ decay products overlap in space-time, 
interference between identical bosons originating from different \PW's can be
expected~\cite{Lonn95,
               Sark-thesis,Lonn98}.
This \interW\ BE interference may provide a laboratory to measure the 
space-time development of this overlap.  
Moreover, this effect, like 
colour reconnection~\cite{Gustaf88,Sjos94, Gustaf94, Lonn96,
                          Sark-thesis,
                          Khoze99},
can be a source of bias in the determination of the \PW\ mass in the four jet channel.
Recent model predictions~\cite{Lonn95,*Ball9601,*Kart97,*Jadach97,*Fial97,*Fial98,
                  Sark-thesis,
                  Lonn98},
as well as recent experimental 
results~\cite{delphi-145},
are still contradictory. 

The main question we address in this paper is, therefore, whether \interW\ BE interference exists.
However, we also examine all BE interference, \intraW\ as well as \interW,
and make a comparison with that observed in hadronic \PZ\ decays, with and without the contribution of
$\PZ\rightarrow\Pqb\Paqb$ decays.

\section*{Analysis Method}

%\subsection*{The \BE\ Correlation Function}

\BE\ interference manifests itself through correlations between identical bosons at small
four-momentum difference.
Correlations between two particles are described by the ratio of the two particle number density,
$\rho_2(p_1,p_2)$, to the product of the two single particle number densities, $\rho_1(p_1)\rho_1(p_2)$.
Since we are only interested in \BE\ correlations (BEC) here, 
the product of single particle densities is replaced by $\rho_0(p_1,p_2)$,
the two particle density that would occur in the absence of \BE\ interference, 
resulting in the BE correlation function 
\begin{equation}   \label{refie}
   R_2(p_1,p_2)  =  \frac{\rho_2(p_1,p_2)}{\rho_0(p_1,p_2)}  \quad.
\end{equation}
For identical bosons, $R_2-1$ is related to the space-time particle density through a Fourier
transformation~\cite{GGLP60,Boal90}.

Since we shall consider only pion pairs, the mass of the particles is fixed 
and the correlation function is defined in six-dimensional momentum space.
Since \BE\ correlations are largest at small four-momentum difference,
$Q\equiv \sqrt{-(p_1-p_2)^2}$, we parametrize $R_2$ in terms of this single variable.
While this is an oversimplification, as recent two- and three-dimensional analyses have 
shown~\cite{l3-174},
lack of statistics prevents such multi-dimensional analyses here.

%\subsection*{Inter-{\kern -0.1em\boldmath\PW}  \BE\ Correlations}

The following method~\cite{CWK99} is used to study \interW\  BEC.
If the two \PW's decay independently, the two particle density in fully hadronic \WW\ events,
$\rho_2^{\WW}$, is given by
\begin{equation}     \label{eq1a}
   \rho_2^{\WW}(p_{1},p_{2})  = \rho_2^{\PWp}(p_{1},p_{2}) + \rho_2^{\PWm}(p_{1},p_{2})
                              + \rho_1^{\PWp}(p_{1})\,\rho_1^{\PWm}(p_{2})
                              + \rho_1^{\PWm}(p_{1})\,\rho_1^{\PWp}(p_{2})  \quad,
\end{equation}
where the superscript, \PWp\ or \PWm, indicates the \PW\ which produced the particles.
Assuming that the densities for \PWp\ and \PWm\ are the same, eq.\,(\ref{eq1a}) becomes
\begin{equation}     \label{eq1b}
   \rho_2^{\WW}(p_{1},p_{2}) = 2\rho_2^{\PW}(p_{1},p_{2})
                             + 2\rho_1^{\PW}(p_{1})\,\rho_1^{\PW}(p_{2})   \quad.
\end{equation}
The terms $\rho_2^{\PW\PW}$ and $\rho_2^\PW$ of eq.\,(\ref{eq1b})
are measured in the fully hadronic \PW\PW\ and the semi-hadronic events, respectively.
To measure the product of the single particle densities, we use the two particle
density $\rho^{\PW\PW}_{\mix}(p_1,p_2)$ obtained by pairing
particles originating from two different semi-hadronic \WW\ events 
($\PWp\PWm\rightarrow\ell\Pgn\Pq\Paq$), since
by construction these pairs of particles are uncorrelated.

The hypothesis that the two \PW's decay independently can be tested using eq.\,(\ref{eq1b}). 
In particular, we write eq.\,(\ref{eq1b}) in terms of $Q$ and use the test statistics
\begin{equation}     \label{del}   
   \Delta \rho(Q) = \rho_2^{\PW\PW}(Q) - 2\rho_2^{\PW}(Q) - 2\rho^{\PW\PW}_{\mix}(Q)
\end{equation}
and
\begin{equation}     \label{eq3}
   D(Q)        = \frac{\rho_2^{\WW}(Q)}{2\rho_2^{\PW}(Q)+2\rho^{\WW}_{\mix}(Q)} \quad.
\end{equation}
The advantage of this method is that it
gives access to the \interW\  correlations directly from the experimental data;
there is no need for normalization by a Monte Carlo (\MC) model.

It is possible that the event mixing procedure introduces artificial distortions and that it does not fully
account for some non-BE correlations or some detector effects.  To diminish the effect of such
inadequacies and to be able to compare more directly to other experiments,
we also use the double ratio
\begin{equation}   \label{eq4}
   D'(Q) = \frac{D(Q)}{D_{\MC,\,\noBE}(Q)} \quad,
\end{equation}
where $D_{\MC,\,\noBE}$ is derived from a Monte Carlo sample with no BEC,
or at least without \interW\ BEC.

In the absence of \interW\ correlations, $\Delta\rho=0$ and $D=D'=1$.
To study BEC, we examine these relations 
for small values of $Q$, for like-sign particles.
To judge the influence of other correlations on these quantities, 
we examine them also for unlike-sign particles and in Monte Carlo models.

\section*{Data Selection}
The data used in this analysis were collected in 1998 by the
L3 detector~\cite{l3-00}, 
and correspond to an integrated luminosity of about 177 pb$^{-1}$ at a
centre-of-mass energy of $\sqrt{s}\simeq 189\,\GeV$.

To obtain the two \PWp\PWm\ event samples, one fully hadronic
and the other semi-hadronic, we reconstruct the visible final state fermions,
\ie, electrons, muons, $\tau$ jets (corresponding to the visible $\tau$-decay
products) and the hadronic jets corresponding to quarks, and apply the selection
criteria described in Ref.~\cite{evsel},
with the additional requirement for the fully hadronic channel that the neural network output must be
greater than 0.6. 
In total, 1032 semi-hadronic events and 1431 fully hadronic events are selected.

The event generator \KORALW~\cite{koralwvers}
is used to simulate the signal processes.  Within \KORALW\ BEC are simulated
using the so-called BE$_{32}$ or BE$_0$ algorithms
\footnote{The BE$_{32}$ algorithm used the parameter values
\PARJ{92}=1.68 and \PARJ{93}=0.38\,\GeV, whereas BE$_{0}$ used \PARJ{92}=1.50 and \PARJ{93}=0.33\,\GeV.
Both have been tuned to the L3 \PZ\ decay data.}\cite{Lonn98}.  % \cite{PYBOEI}.
For most comparisons we will use BE$_{32}$, since it agrees better with the data.
Where BE$_{0}$ is used it will be explicitly stated.
The BEC are implemented for {\em all}\/ particles, which we refer to as BEA, or only for
particles coming from the {\em same}\/ \PW\ (\intraW\ BEC), which we refer to as BES.
The background processes 
$\:\Pep\Pem\rightarrow\PZ/\gamma\rightarrow \Pq\Paq$,
$\:\Pep\Pem\rightarrow\PZ\PZ$
and $\:\Pep\Pem\rightarrow\PZ\Pep\Pem$
(the last relevant only to the \Pq\Paq\Pe\Pgn\ and \Pq\Paq\Pgt\Pgn\ channels)
are generated using \PYTHIA~\cite{PYTHIA61} with BE$_{0}$.
The generated events are passed through the L3 detector simulation
program~\cite{detsim}, reconstructed  and subjected to the same \WW\ selection criteria as the data.

MC studies using the above generators
show that the selection efficiency for fully hadronic events changes by less
than 0.5\% when BEC (\intraW, or both \intraW\ and \interW) are included.
The efficiencies for the channels
$\Pq\Paq\ell\Pgn$, ($\ell=\Pe,\Pgm,\Pgt$) and $\Pq\Paq\Pq\Paq$
% are found to be 81.5\%, 78.5\%, 49.4\% and 87.2\%, respectively.
  are found to be 82.9\%, 76.7\%, 50.6\% and 87.2\%, respectively. % -- from WW cross section draft
The fractions of background for these channels are 
%        4.8\%, 4.1\%, 12.2\% and 18.6\%,
         4.1\%, 4.1\%, 12.7\% and 18.6\%,     % -- from WW cross section draft
respectively.

The BEC study is based on charged particle information from the central tracker.
Charged tracks are required to have at least 35 (of 62 possible) hits,
and the number of wires from the first to the last hit is required to be at least 45. 
The distance of closest approach (projected onto the transverse plane) of a track to the nominal
interaction vertex is required to be less than 7.5 mm. The transverse momentum of a track
must be greater than 100 \MeV.
After the track selection, there are 287k pairs of like-sign particles in the
fully hadronic channel and 55k pairs % of like-sign particles
in the semi-hadronic channel.

\begin{figure}
\begin{center}
\begin{tabular}{c@{\hspace{-2.0cm}}c}
\hspace{-.5cm}
\epsfig{figure=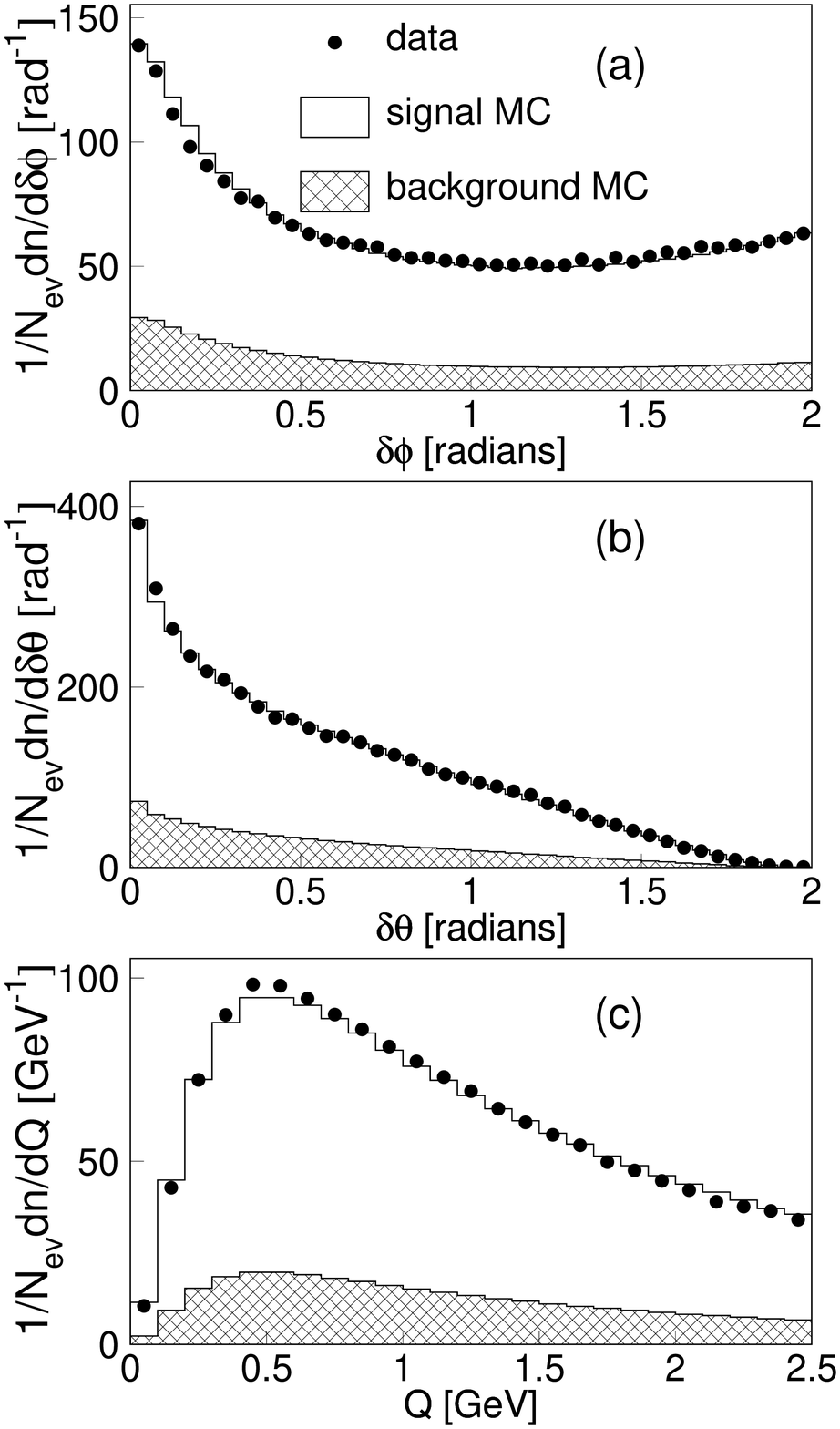, width=.56\linewidth} &
\epsfig{figure=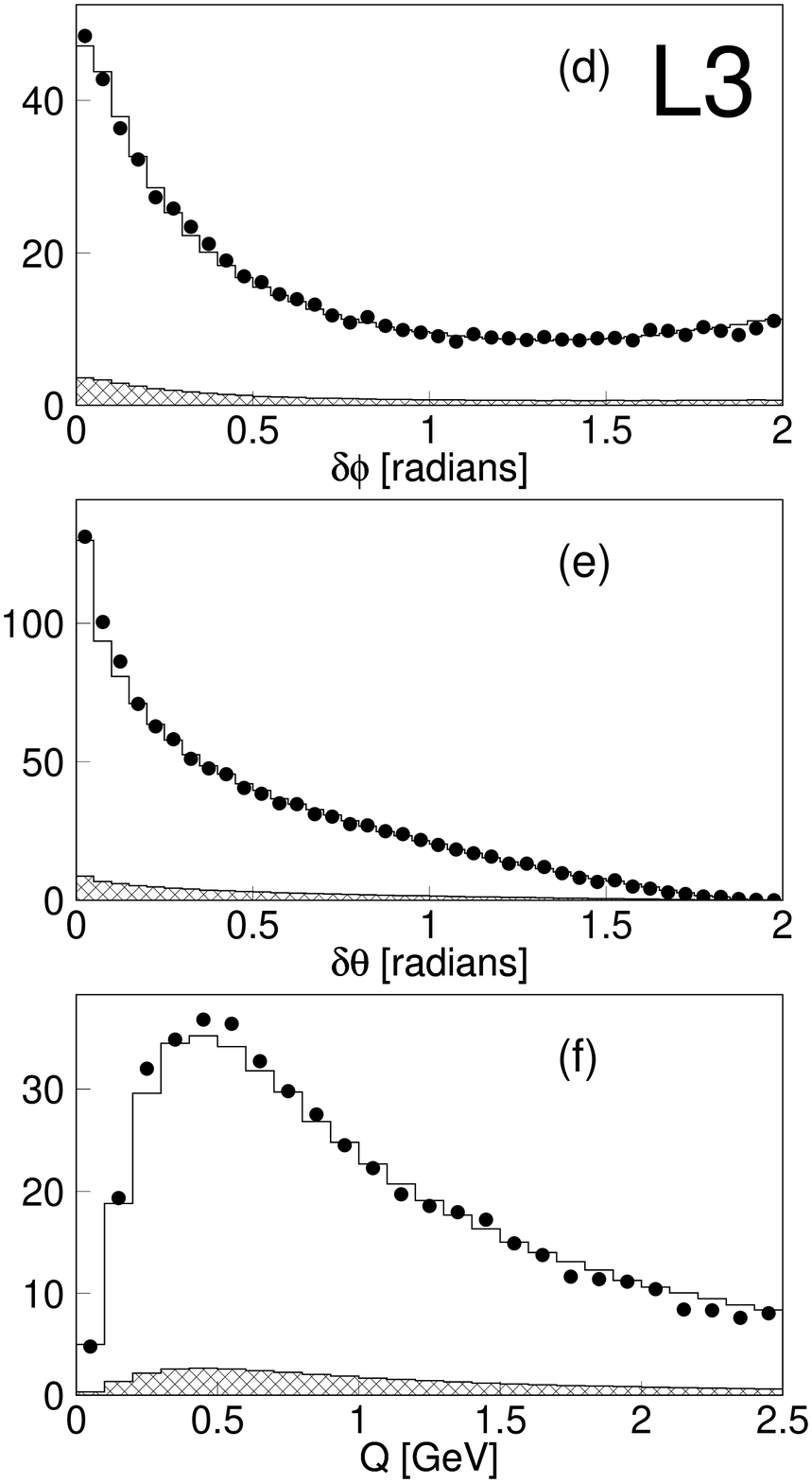, width=.56\linewidth}
\end{tabular}
\icaption{Distributions of 
(a, d) $\delta\phi$, the difference in azimuthal angle of pairs of tracks,
(b, e) $\delta\theta$, the difference in polar angle of pairs of tracks, and
(c, f) $Q$, the four-momentum difference,
for the fully hadronic \WW\ events (a--c) and
for the semi-hadronic \WW\ events (d--f).
Only like-sign pairs of tracks are considered.
The points are the uncorrected data,
the open histograms are the expectation of \KORALW\ with \intraW\ BEC plus background.
The shaded histogram is the background expectation.
\label{figdatsim} }
\end{center}
\end{figure}

With this selection, reasonable agreement is obtained between
the data and the \MC\ simulation for the distributions of $Q$ and  the difference in azimuthal,
as well as polar, angle with respect to the beam, for pairs of like-sign tracks,
in both the fully hadronic and the semi-hadronic channels. 
This is shown in Fig.\,\ref{figdatsim},
where the raw data are compared to simulated \KORALW\ with BES and background events.
Similar agreement is observed when BEA is used.

\section*{Measurement of \boldmath $R_2$}

We first measure the BE correlation function, $R_2$, for like-sign charged pion pairs using two choices of
reference sample, \ie, the sample from which $\rho_0$ is determined.
The first choice uses a Monte Carlo model without BEC:
\begin{equation}                    \label{refer0} 
  \rho_{0}(\pm,\pm)=\rho_2(\pm,\pm)_{\MC,\,\noBE}   \quad.
\end{equation}
The second choice uses unlike-sign particle pairs from the experimental events. 
A major drawback of this method is that the correlation function is affected by the presence
of dynamical correlations, such as the decay of resonances.
To compensate for this, the density for unlike-sign pairs is multiplied by the ratio
of the densities for like- and unlike-sign pairs determined from a Monte Carlo model without BEC:
\begin{equation}                    \label{refer1}
  \rho_{0}(\pm,\pm) = \rho_2(+,-) \cdot \left[\frac{\rho_2(\pm,\pm)}{\rho_2(+,-)}\right]_{\MC,\,\noBE} \quad. 
\end{equation}

In both cases we need to correct the correlation function, $R_2$, for detector resolution, acceptance,
efficiency and for particle misidentification.
For this we use a multiplicative factor derived from Monte Carlo studies. 
Since we do not perform explicit hadron identification, this factor is
given by the ratio of $\rho_2(\pm,\pm)$ and $\rho_2(\pm,\pm)/\rho_2(+,-)$, respectively,
found from Monte Carlo, for {\em pions}\/ at generator level
to that found using {\em all particles}\/ after full detector simulation, reconstruction and selection.
Thus, using eqs.\,(\ref{refer0}) and (\ref{refer1}) leads, respectively, to
\begin{equation}          \label{funct0}
  R_2 = \left[\frac{\rho_2(\pm,\pm)_{\data}}{\rho_2(\pm,\pm)_{\MC,\,\noBE}}\right] \cdot
        \left[\frac{\rho_2(\pm,\pm)_{\gen}}{\rho_2(\pm,\pm)_{\det}}\right]_{\MC} 
\end{equation}
and
\begin{equation}          \label{funct1}
  R_2 = \left[\frac{\rho_2(\pm,\pm)}{\rho_2(+,-)}\right]_{\data} \cdot
         \left[\frac{\rho_2(+,-)}{\rho_2(\pm,\pm)}\right]_{\MC,\,\noBE} \cdot
         \left[\frac{\rho_2(\pm,\pm)_{\gen}}{\rho_2(\pm,\pm)_{\det}}  \cdot
               \frac{\rho_2(+,-)_{\det}}{\rho_2(+,-)_{\gen}}\right]_{\MC}       \quad.
\end{equation}

Background is taken into account by replacing $\rho_{2\,\mathrm{data}}$ in the above equations by
\begin{equation}
  \rho_{2\,\data-\mathrm{bg}} = \frac{1}{\mathcal{P} N_\mathrm{ev}}
                               \left( \frac{\dd{n}}{\dd{Q}}
                                    - \frac{\dd{n_\mathrm{bg}}}{\dd{Q}} \right) \quad,
\end{equation}
where $\mathcal{P}$ is the purity of the selection, $n$ is the number of pairs of tracks in the
$N_\mathrm{ev}$ data events, and $n_\mathrm{bg}$ is the number of pairs
of tracks corresponding to $(1-\mathcal{P})N_\mathrm{ev}$ background events.
The background is estimated using Monte Carlo.
In determining $R_2$ using eq.\,(\ref{funct0}), we use \KORALW\ without BEC as the reference sample.
For the detector correction, BES with the BE$_{32}$ algorithm is used for both eqs.\,(\ref{funct0}) and
(\ref{funct1}).

\begin{figure}
\begin{center}
\epsfig{figure=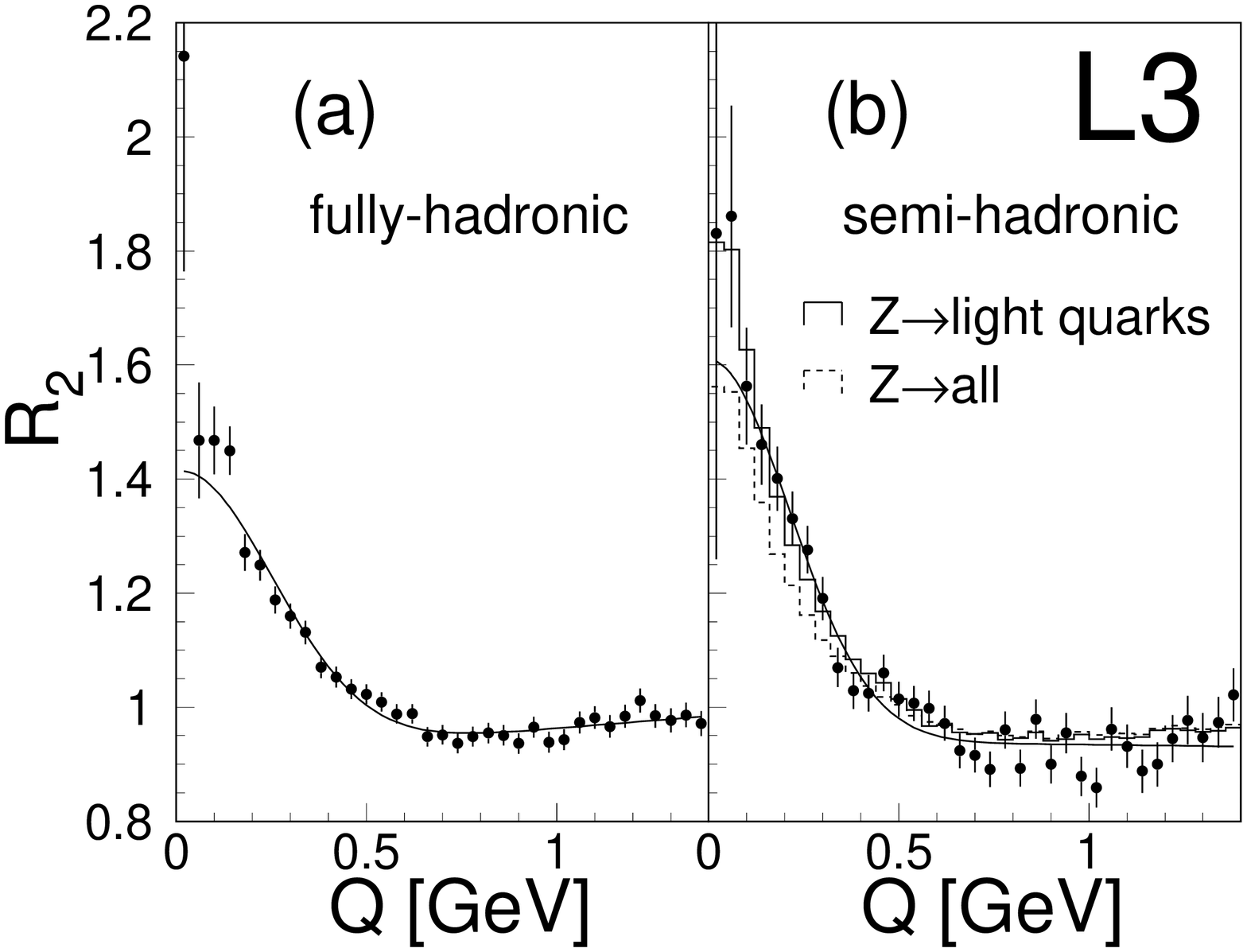, width=1.04\linewidth}
\icaption{The \BE\ correlation function $R_{2}$, eq.\,(\ref{funct0}), for
(a) the fully-hadronic \WW\ events, and (b) the semi-hadronic \WW\ events.
In (b) the full histogram is for the light-quark \PZ\ decay sample and the dashed histogram is
for a sample containing all hadronic \PZ\ decays.
Also shown are the fits of eq.\,(\ref{fitg}) to the \WW\ data.
\label{figr2} }
\end{center}
\end{figure}

Fig.\,\ref{figr2} shows the correlation function, eq.\,(\ref{funct0}), for the fully hadronic and for the
semi-hadronic \WW\ events.
We parametrize the \BE\ enhancement at low $Q$ values by
\begin{equation}   \label{fitg}
  R_{2}(Q) = \gamma (1+\delta Q) (1+\lambda \exp(-R^2 Q^2)) \quad,
\end{equation}
where $\gamma$ is an overall normalization factor, the term $(1+\delta Q)$ takes into account
possible long-range momentum correlations, $\lambda$ measures the strength of the BE correlations
and $R$ is related to the source size in space-time.
The results of the fits of eq.\,(\ref{fitg}) are also shown in Fig.\,\ref{figr2}.

The fit results for both choices of reference sample, eqs.~(\ref{refer0}) and (\ref{refer1}),
are given in Table~\ref{table1}.
The statistical error includes bin-to-bin correlations.           
These are estimated from 100 sets of \PWp\PWm\ BES events generated by \PYTHIA,
each with the same statistics as the data.
The variation of $\gamma$, $\lambda$, $R$ and $\delta$ from their average 
values was determined for the fully- and semi-hadronic \WW\ events. 
The ratio of the Gaussian width to the average fit error was found to be between 1.01 and 1.61.
For each parameter, the corresponding ratio is used to scale the original statistical error.
MC studies show that this ratio hardly depends on $Q$,
which justifies this method to correct for bin-to-bin correlations.

The systematic uncertainty is computed by varying the track and event selections.
Both stronger and weaker cuts are applied to the tracks, slightly different
event selections are made, and the background fractions are varied.
The influence of the choice of the Monte Carlo used for the reference sample
and for the correction factor is also taken into account. 
Part of the systematic uncertainty comes from the choice of the fit range. 
The large systematic uncertainty on $\lambda$ in the fully hadronic
channel is mainly due to the difference of including
or not including \interW\ BEC in the \MC\ for the correction factor.
Using \interW\ BEC in the correction factor increases the measured value of~$\lambda$.

BEC are observed ($\lambda > 0$) in both fully hadronic and semi-hadronic \WW\ events.
The values of $\lambda$ are higher for the semi-hadronic than for the fully hadronic channel,
%suggesting a suppression of \interW\ BEC \cite{CWK99},
but the difference is only about two standard deviations (statistical error only)
for each choice of reference sample.
If true, this effect would indicate a suppression of \interW\ BEC \cite{CWK99}, which we study in detail in
the following section.

Since, apart from the quark flavour, hadronic \PW\ and \PZ\ decays are expected to be similar,
we also analyze a high statistics hadronic \PZ\ decay sample, collected
by the L3 detector in 1994 at $\sqrt{s}\simeq 91.2 \GeV$.
Since b quarks are greatly suppressed in \PW\ decays,
a \Pqb-tagging procedure~\cite{l3-127,*dominguez_thesis}        
is used to reduce the \Pqb\Paqb\ fraction in \PZ\ decays, from 22\% to 3\%. 
The BE correlation function, eq.\,(\ref{funct0}), of the resulting 180k \PZ\ events
is plotted in Fig.\,\ref{figr2}b as a full histogram.
As expected, good agreement is observed between this histogram and the correlation function
of the semi-hadronic \WW\ events. 
When b quark decays of the \PZ\ are not removed from the sample,
a depletion of the correlation function at small $Q$ is observed and
a clear discrepancy exists with the \PW\ data, as the dashed histogram in Fig.\,\ref{figr2}b shows.

\section*{Measurement of Inter-{\kern -0.1em\boldmath\PW} \BE\ Correlations}   \label{interwbec}

\subsection*{The Event Mixing Procedure}

To compute the test statistics, eqs. (\ref{del}) and (\ref{eq3}), we need to construct the
two particle density $\rho^{\PW\PW}_{\mix}$.
This is done by combining pairs of
semi-hadronic events having oppositely charged hadronically decaying \PW's.
Particles identified as decay products of the leptonically decaying \PW's are discarded.
Then the particles from one of the events are rotated 
so that the \PW's are approximately back-to-back.
Since real fully hadronic \WW\ events have a small longitudinal energy imbalance
that we ascribe to initial state radiation and since experimental resolution
leads to both transverse and longitudinal energy imbalance, we do not force the \PW's to be exactly
back-to-back. We introduce an extra momentum, $\vec{p}_{\mathrm{extra}}$, Gaussian distributed in all three
components and impose $\vec{p}_{\mathrm{extra}}+\vec{W}_{1}=-\vec{W}_{2}$,
where $\vec{W}_{1,2}$ are the momenta of the two W's.
For the longitudinal component the Gaussian has mean 0 and standard deviation 7.9 \GeV, while for the
transverse components the mean is randomly chosen as $\pm0.5$ \GeV\ and the standard deviation is 1.4 \GeV.
These values were chosen to obtain reasonable agreement 
between the energy imbalance distributions of fully hadronic and mixed events.

In addition, we impose the following cuts which are related to the pre-selection of fully hadronic
\WW\ events~\cite{evsel}.
We demand
that the sphericity be larger than 0.045,
that the total visible energy  be larger than $0.7\sqrt{s}$,
that the number of particles identified with the calorimeter (the cluster multiplicity) be larger than 30, 
that the ratio of the total longitudinal energy imbalance to the visible energy be smaller than 0.25,
and that the $y_\mathrm{cut}$ value at which the event changes from a 3- to a 4-jet topology, $y_{34}$, be
larger than 0.001. After forcing the event into 2 jets with the Durham clustering algorithm \cite{durham},
the average of the jet masses is required to be larger than 30 \GeV.
After forcing the event into 4 jets with the Durham clustering algorithm,
we assign two pairs of jets to each of the two W's by first rejecting the combination with the smallest
dijet mass and then accepting the combination with the smallest difference between the two dijet masses 
% the remaining two pairing possibilities 
(best pairing).
We then demand
that the difference between the two \PW\ masses be less than 70 \GeV,
that the smallest angle between any two jets be larger than 0.28 radians,
and that the average of the two smallest angles between two jets from different \PW's 
   be larger than 0.6 radians.
These cuts reject only approximately 1\% of the events and do not change the $Q$-distribution.
The final selection of fully hadronic events uses a cut at 0.6 on the output of a neural
network~\cite{evsel}. 
This cut is also applied to the mixed events, rejecting 7\%.    %   7.1\%.

\begin{figure}
\begin{center}
\epsfig{figure=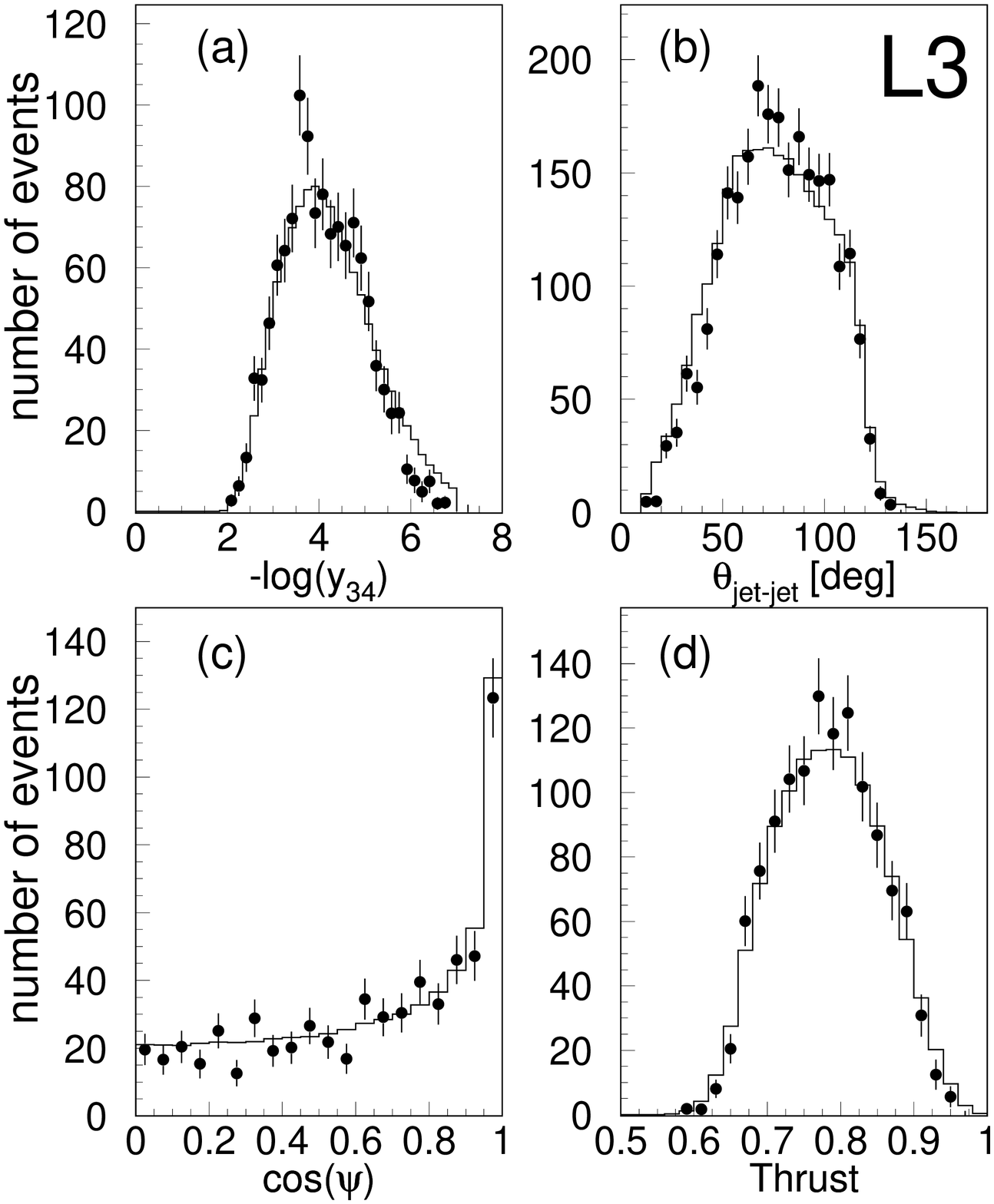, width=1.05\linewidth}
\icaption{Comparison of uncorrected distributions for fully hadronic events after background
subtraction (points) and mixed events (histograms):
(a) $-\log y_{34}$;
(b) the two smallest angles between jets of different \PW's, after jet finding and best pairing;
(c) the cosine of the angle $\psi$ between the decay planes of the two \PW's, after jet finding and best pairing; 
and (d) the event thrust.
\label{fig3_0} }
\end{center}
\end{figure}

We have checked the mixing procedure by comparing the distributions
and quantities of a large number of variables between mixed events and 
fully hadronic \WW\ events, including event shape variables, track and cluster multiplicities,
and variables related to the W such as mass, energy and orientation.
In general, good agreement is found. Typical examples are shown in Fig.\,\ref{fig3_0}.

\subsection*{Results}

\begin{figure}
\begin{center}
\epsfig{figure=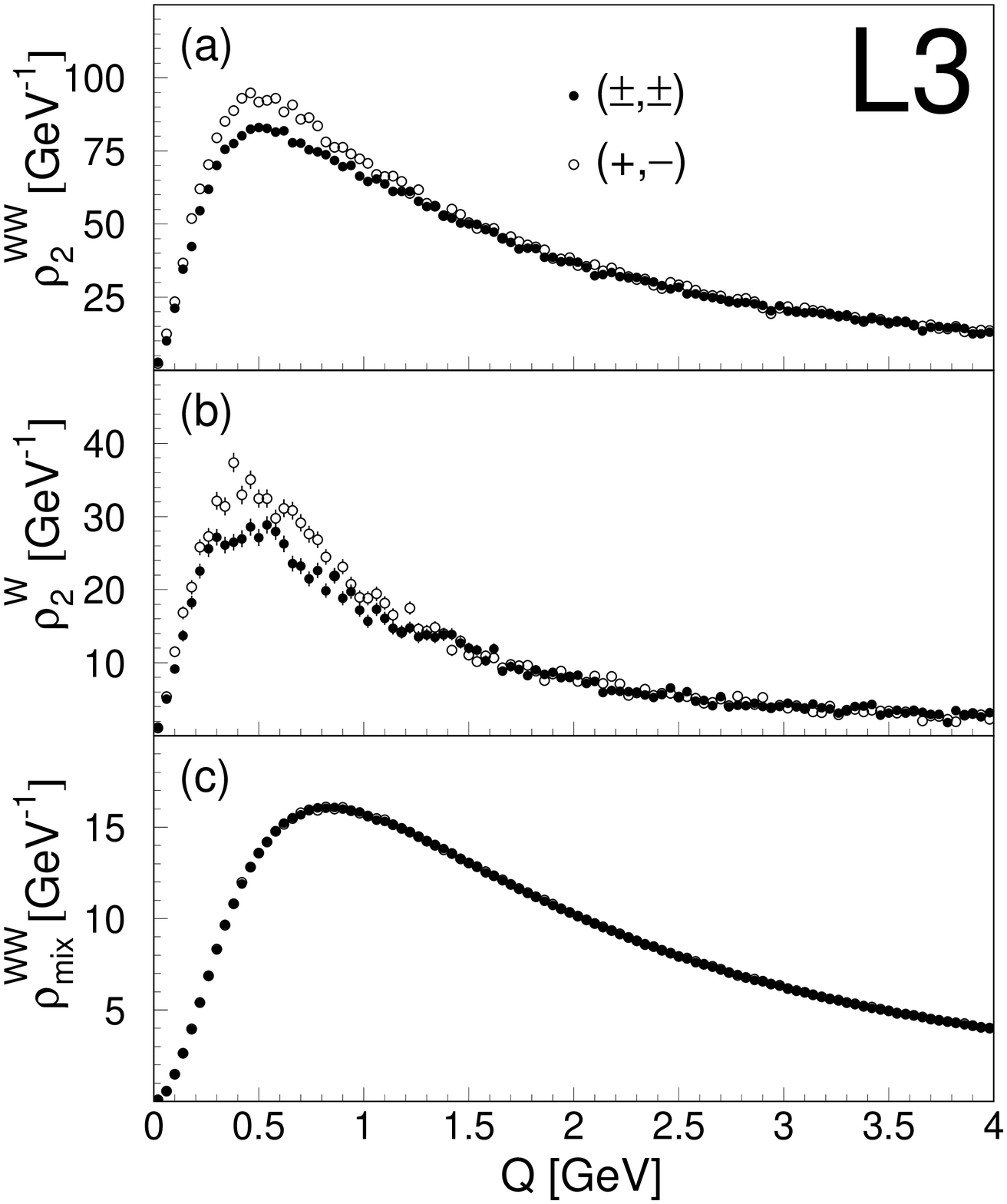, height=19cm,width=1.1\linewidth}
\icaption{Distributions for uncorrected data
of (a) $\rho_2^{\WW}$, (b) $\rho_2^{\PW}$ and (c) $\rho^{\WW}_\mathrm{mix}$
for pairs of like-sign charged particles 
and pairs of unlike-sign charged particles.
\label{fig3} }
\end{center}
\end{figure}

Fig.\,\ref{fig3} shows the distributions of the three terms in the right-hand side of
eq.\,(\ref{del}) for the data, not corrected for detector effects,
but after background subtraction. 
At low values of $Q$  we observe more pairs of unlike-sign particles
than pairs of like-sign particles, both in the two-particle densities
for fully hadronic (Fig.\,\ref{fig3}a) and semi-hadronic (Fig.\,\ref{fig3}b) events.
Furthermore, we observe that $\rho^{\WW}_{\mix}(\pm,\pm)$
and $\rho^{\WW}_{\mix}(+,-)$ coincide (Fig.\,\ref{fig3}c).

\begin{figure}
\begin{center}
\epsfig{figure=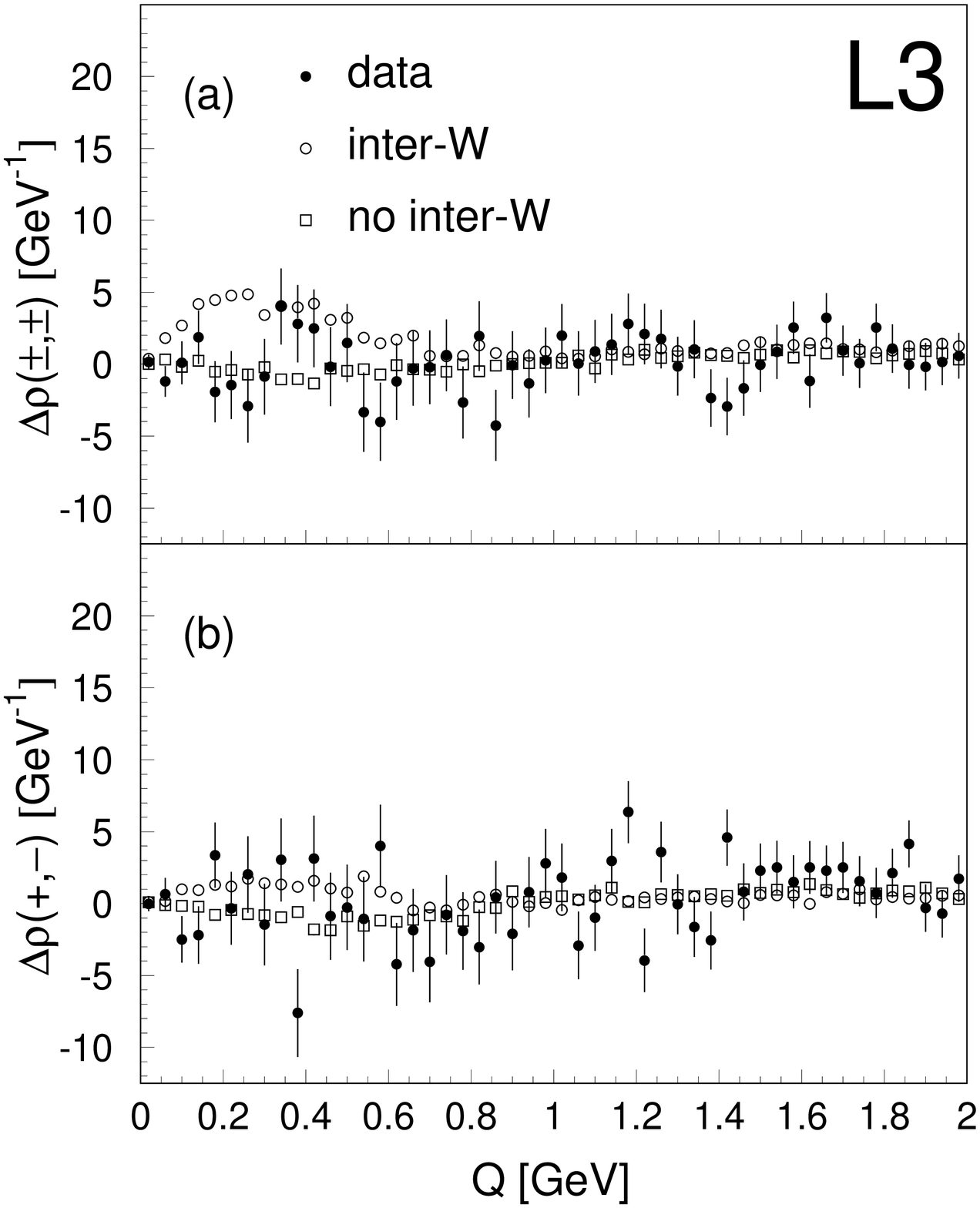, width=1.05\linewidth}
\icaption{Distributions for uncorrected data of 
(a) $\Delta \rho(\pm,\pm)$  and
(b) $\Delta \rho(+,-)$.
Also shown are the Monte Carlo predictions of \KORALW\ (at the detector level) 
with BEA (\interW) and BES (no \interW).
\label{fig4}  }
\end{center}
\end{figure}

From these distributions,
we compute $\Delta\rho$ for like-sign and unlike-sign particle pairs, eq.\,(\ref{del}).
The resulting raw data distributions are shown in Fig.\,\ref{fig4}. Also shown are the
predictions of \KORALW\ after full detector simulation, reconstruction and selection.
Both the BEA and BES scenarios are shown.

The BEA scenario using BE$_{32}$
shows an enhancement in the $\Delta\rho$ distribution for
like-sign pairs (Fig.\,\ref{fig4}a), but also a small enhancement for unlike-sign pairs (Fig.\,\ref{fig4}b).
The effect for unlike-sign pairs is larger if BE$_0$ is used.
These implementations of BEC clearly affect both the like- and unlike-sign particle spectra. 
From Fig.\,\ref{fig4}a it is clear that only
the BES scenario describes the $\Delta\rho(\pm,\pm)$ distribution,
while the BEA scenario is disfavoured. 
In particular, considering $Q$-values up to 0.6 \GeV, the
confidence level (CL) for the BES and BEA scenarios are, respectively, 84\% and 0.8\%. The calculations
of the confidence levels are based on statistical errors including bin-to-bin correlations.

\begin{figure}
\begin{center}
\epsfig{figure=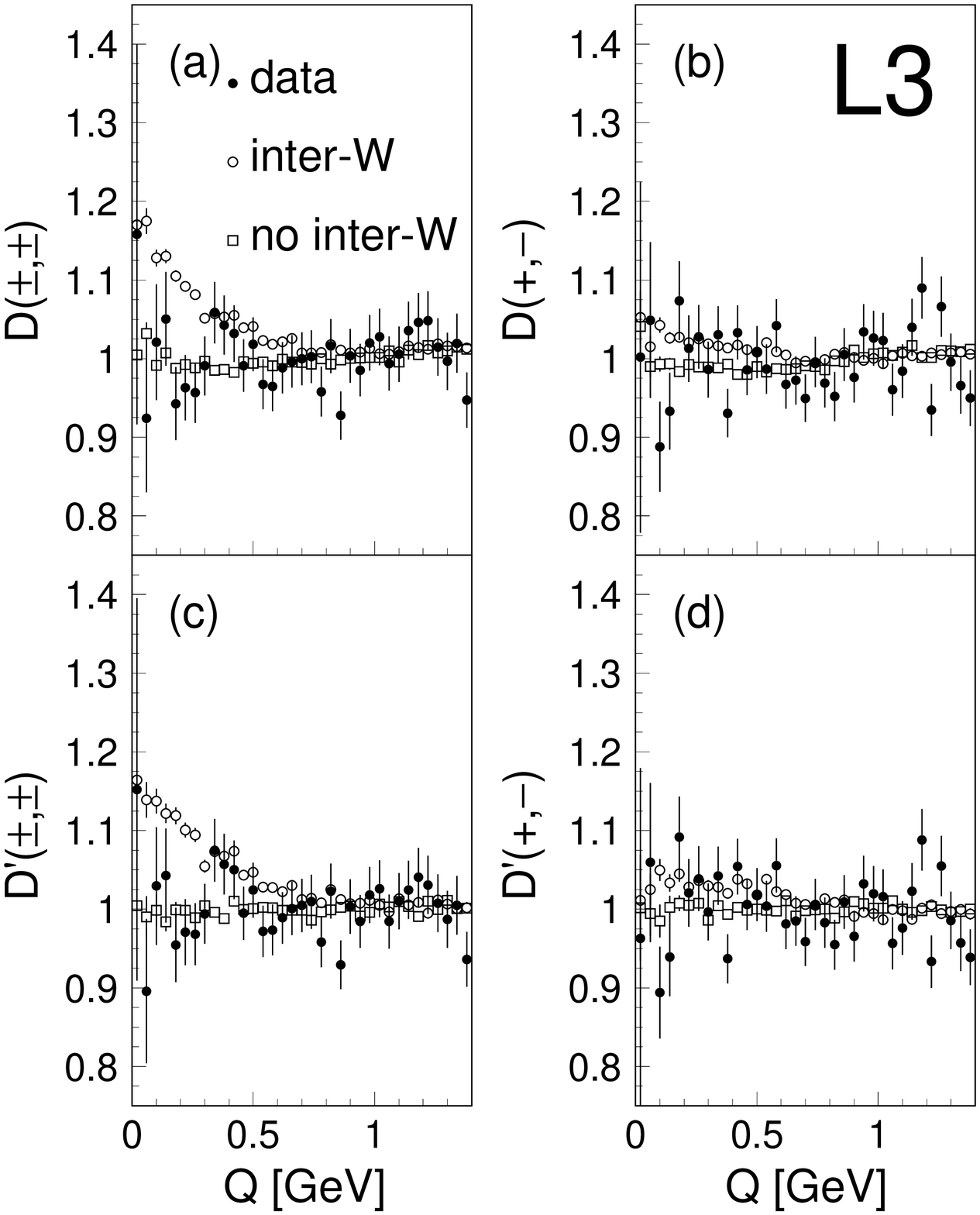, width=1.1\linewidth}
\icaption{Distributions for uncorrected data of 
(a) $D(\pm,\pm)$, 
(b) $D(+,-)$, 
(c) $D'(\pm,\pm)$ and 
(d) $D'(+,-)$. 
Also shown are the predictions of \KORALW\ (at the detector level) with 
BEA (\interW) and BES (no \interW).
\label{fig5}  }
\end{center}
\end{figure}

In Fig.\,\ref{fig5} we show the
distributions of $D$ and $D'$ for like-sign and unlike-sign
particle pairs, eqs.\,(\ref{eq3}) and~(\ref{eq4}), for the raw data.
For the double ratio $D'$ we use the BES scenario of \KORALW\ as the reference sample.
Also shown in the figure are the predictions of \KORALW\   % (on the detector level Monte Carlo) 
for the scenarios BEA and BES.
Again, it is clear that the BES scenario of \KORALW\ describes the data, 
CL=87\% for both $D(\pm,\pm)$ and $D'(\pm,\pm)$,
while the BEA scenario is disfavoured, CL=0.5\% for both $D(\pm,\pm)$ and $D'(\pm,\pm)$.
When BE$_0$ is used instead of BE$_{32}$,
the BEA scenario is even more strongly disfavoured: CL=0.08\% for both $D(\pm,\pm)$ and
$D'(\pm,\pm)$.   %  0.07\% for %D%,  0.09\% for $D'$
Note that the $D'$ distributions are by definition equal to unity (apart from
statistical fluctuations) when \KORALW\ without \interW\ BEC is used.
Note also that $D$ is already close to unity for BES, so that
the difference between $D$ and $D'$ is small, which supports the validity of the mixing procedure.

To estimate the strength of \interW\ BEC, the $D'(\pm,\pm)$ distribution is fitted (from 0 to 1.4 \GeV)
by the following function 
\begin{equation}
  D'(Q) = (1+\epsilon Q)\left( 1+ \Lambda \exp(-k^2 Q^2)\right) \quad,
\end{equation}
where $\epsilon$, $\Lambda$ and $k$ are the fit parameters. 
%The parameter $\Lambda$ measures the strength of inter-W BEC.
%
The result of the fit for the strength of \interW\ BEC is
$$  \Lambda=0.001\pm 0.026\pm 0.015 \quad,   $$
where the first error is statistical and the second systematic.
The statistical error has been multiplied by 1.49 to account for  bin-to-bin correlations, in the same way
as described in the previous section.  This value of $\Lambda$ is consistent with zero, \ie, with no
\interW\ BEC. A similar fit was performed for the \KORALW\ BEA distribution, resulting in $\Lambda=0.127\pm0.007$
(statistical error only).  The data disagree with this value by more than 4 standard deviations.  % 4.1 std dev

The systematic uncertainty on~$\Lambda$ is the sum in quadrature of the contributions listed in
Table~\ref{table2}. The amount of background was varied by $\pm10\%$.
%The choice of Monte Carlo was varied using \PYTHIA\ as well as \KORALW\ and with no BEC at all, as well as
The choice of Monte Carlo was varied using \PYTHIA\ and \KORALW, both with no BEC at all as well as
with only \intraW\ BEC.
Also the effect of various models of colour reconnection\footnote{The so-called SKI, SKII, SKII$^\prime$
\cite{Sjos94} and GH \cite{Gustaf94} models, as implemented in PYTHIA, were used.} (CR) was included.
A change in the fit range ($\pm0.4\,\GeV$), a change in the bin size (from 40 to 80\,\MeV)
and a change in the parametrization (removing the factor $(1+\epsilon Q)$ from the fit) 
also give contributions to the systematic uncertainty. 
Furthermore, the track and event selections were varied.

In the mixing procedure we allow a semi-hadronic \WW\
event to be combined with all possible other semi-hadronic  \WW\ events. 
To be sure that this does not introduce a bias, 
the analysis was repeated for a mixed sample where every semi-hadronic %  \WW\
event was used only once.
The influence of the mixing procedure was also studied by not only combining oppositely charged \PW's,
but also like-sign \PW's.
The influence of the extra momentum $\vec{p}_{\mathrm{extra}}$, used 
in the event mixing, is also included as a systematic effect.
The RMS of the systematic uncertainties due to these three changes in the mixing procedure is the
systematic uncertainty
listed in Table~\ref{table2}.
The influence of the cut on the neural network output for the mixed events was investigated by removing the
cut.

Furthermore, the effect of uncertainties in the energy calibration of the calorimeters was studied.
Finally, we studied the influence of the \Pq\Paq\Pgt\Pgn\ channel.
Since this channel is the most difficult to identify, and therefore
has relatively high background and low efficiency, we repeated the analysis without it.

\begin{figure}
\begin{center}
\epsfig{figure=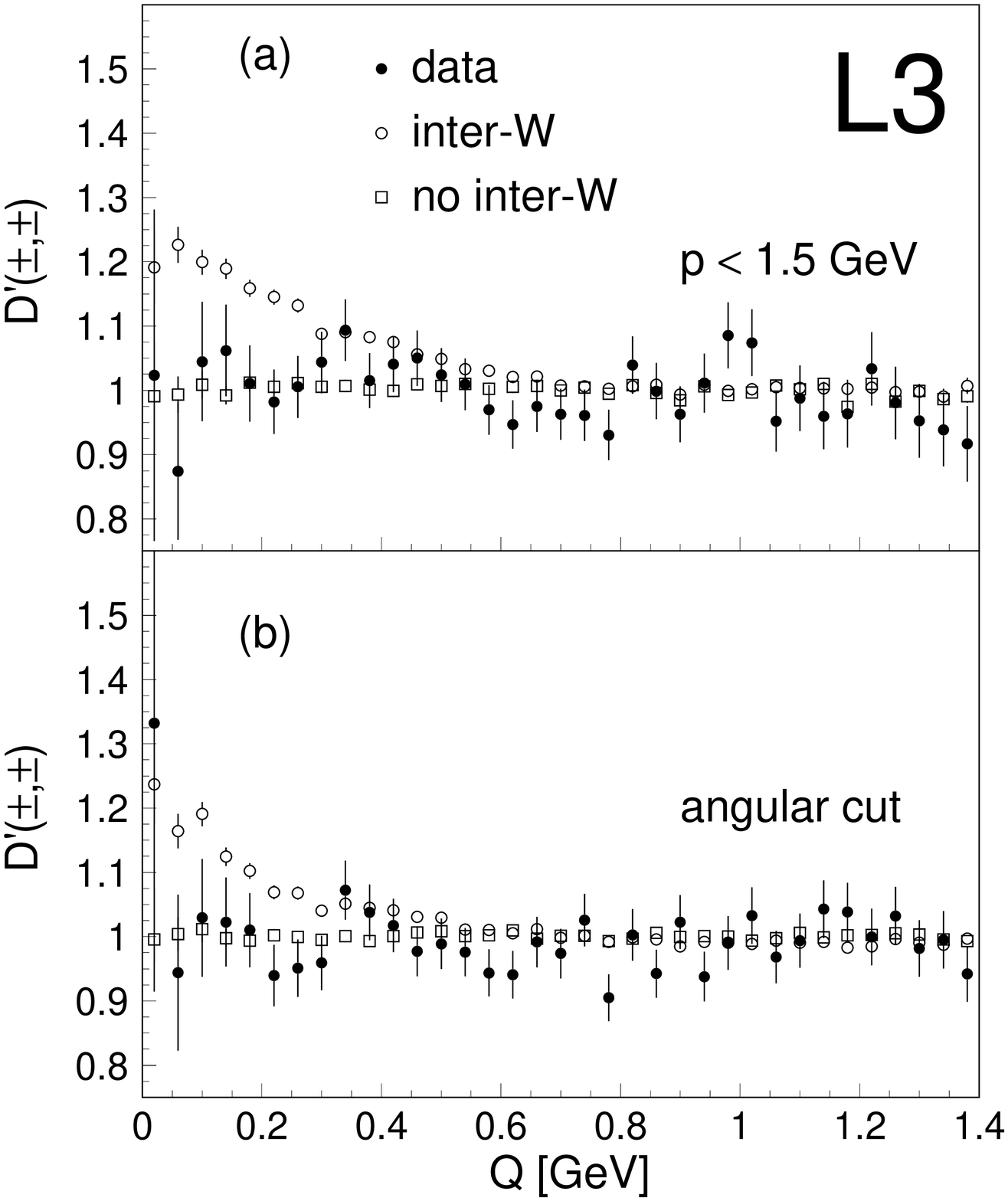, width=1.05\linewidth}
\icaption{Distributions for uncorrected data of $D'(\pm,\pm)$ where
(a) only low momentum tracks are used and
(b) a cut is made on the average angle of the two smallest angles between jets of different \PW's.
Also shown are the predictions of \KORALW\ (at the detector level) with 
BEA (\interW) and BES (no \interW).
\label{fig6}  }
\end{center}
\end{figure}

To make the analysis possibly more sensitive to \interW\ BEC, we repeated the analysis twice
using different selections to increase the overlap of the \PWp\ and \PWm\ decay products.
Since BEC occur mainly among soft particles and the overlap is expected to be larger for these particles
than for high-momentum ones, we first repeated the analysis using only tracks with momenta smaller than
1.5\,\GeV.
Another way to increase the overlap is to require that jets from different \PW's be close together.
We therefore repeated the analysis requiring that the average of the smallest two  of the four angles
between jets from different \PW's   %  (after jet finding and best pairing)
be less than $75^\circ$. 
This results in a reduction of approximately 60\% in the number of fully hadronic \WW\ events.
Fig.\,\ref{fig6} shows the distributions of $D'(\pm,\pm)$ for these two analyses. 
It is again clear that the
BEA scenario is disfavoured, particularly for the low-momentum selection, while BES describes the data well.
For the low-momentum sample we find CL=1.6\% for BEA and CL=96\% for BES,
and for the sample with the angular cut we find CL= 10\% for BEA and CL=93\% for BES. 
Moreover, we find
$\Lambda= 0.026\pm 0.034$ for the low-momentum sample and
$\Lambda=-0.019\pm 0.029$ for the sample with the angular cut. 
Both values are consistent with zero. 
The errors here are statistical only, including bin-to-bin correlations.

\section*{Conclusion}

{Intra-\kern -0.1em \PW} \BE\ correlations have been found to be similar to those observed in \PZ\ decay
to light quarks. 
An excess at small values of $Q$ in the distributions 
of $\Delta\rho(\pm,\pm)$, $D(\pm,\pm)$ and $D'(\pm,\pm)$ 
is expected from \interW\ BEC, but none is seen.
These distributions agree well with \KORALW\ using BE$_{32}$ when \interW\ BEC are not included,
but not when they are.
We thus find no evidence for BEC between identical pions originating from different \PW's
and disfavour their implementation using the BE$_{32}$ and BE$_0$ algorithms.

%%%%%%%%%%%%%%%%%%%%%%%%%%%%%%%%%%%%%%%%%%%%%%%%%%%%%%%%%%%%%%%%%%%%%%%%%%%%%%%
% Acknowledgements
%%%%%%%%%%%%%%%%%%%%%%%%%%%%%%%%%%%%%%%%%%%%%%%%%%%%%%%%%%%%%%%%%%%%%%%%%%%%%%%
%
\section*{Acknowledgments}

We wish to 
express our gratitude to the CERN accelerator divisions for
the excellent performance of the LEP machine. 
We acknowledge the contributions of the engineers 
and technicians who have participated in the construction 
and maintenance of this experiment.  

%
%%%%%%%%%%%%%%%%%%%%%%%%%%%%%%%%%%%%%%%%%%%%%%%%%%%%%%%%%%%%%%%%%%%%%%%%%%%%%%%
% The author list
%%%%%%%%%%%%%%%%%%%%%%%%%%%%%%%%%%%%%%%%%%%%%%%%%%%%%%%%%%%%%%%%%%%%%%%%%%%%%%%
%
\newpage
\section*{Author List}
\typeout{   }     
\typeout{$Modified: Tue Jun 27 13:17:58 2000 by clare $}
\typeout{!!!!  This should only be used with document option a4p!!!!}
\typeout{   }
%
%
%
%  L A T E X  version!!
%
%
% Make sure that the Lep package has been used!
%\input{Lep.sty}%
%
%\ifx\LepCalled\undefined%
%\typeout{     }%
%\typeout{!!!!!!!!!!!!!!!!!!!!!!!!!!!!!!!!!!!!!!!!!!!!!!!!!!!!!!!!!!!}%
%\typeout{Yikes.  You haven't used the Lep package!}%
%\typeout{Please put \protect\usepackage\protect{Lep\protect} in your preamble,
%         followed by}%
%\typeout{\protect\Lep\protect{1\protect} or \protect\Lep\protect{2\protect}}%
%\typeout{     }%
%\typeout{For now you will get a Lep phase 2 authorlist (may not be right!).}%
%\typeout{!!!!!!!!!!!!!!!!!!!!!!!!!!!!!!!!!!!!!!!!!!!!!!!!!!!!!!!!!!!}%
%\typeout{     }%
%\Lep{2}\fi%

\newcount\tutecount  \tutecount=0
\def\tutenum#1{\global\advance\tutecount by 1 \xdef#1{\the\tutecount}}
\def\tute#1{$^{#1}$}
\tutenum\aachen            % 1
\tutenum\nikhef            % 2
\tutenum\mich              % 3
\tutenum\lapp              % 4
\tutenum\basel             % 5
\tutenum\lsu               % 6
\tutenum\beijing           % 7
\tutenum\berlin            % 8
\tutenum\bologna           % 9 
\tutenum\tata              % 10
\tutenum\ne                % 11
\tutenum\bucharest         % 12
\tutenum\budapest          % 13
\tutenum\mit               % 14 
\tutenum\debrecen          % 15
\tutenum\florence          % 16
\tutenum\cern              % 17 
\tutenum\wl                % 18 
\tutenum\geneva            % 19
\tutenum\hefei             % 20
\tutenum\seft              % 21
\tutenum\lausanne          % 22
\tutenum\lecce             % 23
\tutenum\lyon              % 24
\tutenum\madrid            % 25
\tutenum\milan             % 26
\tutenum\moscow            % 27
\tutenum\naples            % 27
\tutenum\cyprus            % 29
\tutenum\nymegen           % 30
\tutenum\caltech           % 31
\tutenum\perugia           % 32
\tutenum\cmu               % 33
\tutenum\prince            % 34
\tutenum\rome              % 35
\tutenum\peters            % 36
\tutenum\potenza           % 37
\tutenum\salerno           % 38
\tutenum\ucsd              % 39
\tutenum\santiago          % 40
\tutenum\sofia             % 41 
\tutenum\korea             % 42
\tutenum\alabama           % 43
\tutenum\utrecht           % 44
\tutenum\purdue            % 45
\tutenum\psinst            % 46
\tutenum\zeuthen           % 47
\tutenum\eth               % 48
\tutenum\hamburg           % 49
\tutenum\taiwan            % 50
\tutenum\tsinghua          % 51

{
\parskip=0pt
\noindent
{\bf The L3 Collaboration:}
\ifx\selectfont\undefined%  old style font selection
 \baselineskip=10.8pt
 \baselineskip\baselinestretch\baselineskip
 \normalbaselineskip\baselineskip
 \ixpt
\else%                      new style font selection
 \fontsize{9}{10.8pt}\selectfont
\fi
\medskip
\tolerance=10000
\hbadness=5000
\raggedright
\hsize=162truemm\hoffset=0mm
\def\r{\rlap,}
\noindent

M.Acciarri\r\tute\milan\
P.Achard\r\tute\geneva\ 
O.Adriani\r\tute{\florence}\ 
M.Aguilar-Benitez\r\tute\madrid\ 
J.Alcaraz\r\tute\madrid\ 
G.Alemanni\r\tute\lausanne\
J.Allaby\r\tute\cern\
A.Aloisio\r\tute\naples\ 
M.G.Alviggi\r\tute\naples\
G.Ambrosi\r\tute\geneva\
% ...
\newpage
%\rule{\textwidth}{0.4pt}
\begin{list}{A}{\itemsep=0pt plus 0pt minus 0pt\parsep=0pt plus 0pt minus 0pt
                \topsep=0pt plus 0pt minus 0pt}
\item[\aachen]
 I. Physikalisches Institut, RWTH, D-52056 Aachen, FRG$^{\S}$\\
 III. Physikalisches Institut, RWTH, D-52056 Aachen, FRG$^{\S}$
\item[\nikhef] National Institute for High Energy Physics, NIKHEF, 
     and University of Amsterdam, NL-1009 DB Amsterdam, The Netherlands
\item[\mich] University of Michigan, Ann Arbor, MI 48109, USA
\item[\lapp] Laboratoire d'Annecy-le-Vieux de Physique des Particules, 
     LAPP,IN2P3-CNRS, BP 110, F-74941 Annecy-le-Vieux CEDEX, France
\item[\basel] Institute of Physics, University of Basel, CH-4056 Basel,
     Switzerland
\item[\lsu] Louisiana State University, Baton Rouge, LA 70803, USA
\item[\beijing] Institute of High Energy Physics, IHEP, 
  100039 Beijing, China$^{\triangle}$ 
\item[\berlin] Humboldt University, D-10099 Berlin, FRG$^{\S}$
\item[\bologna] University of Bologna and INFN-Sezione di Bologna, 
     I-40126 Bologna, Italy
\item[\tata] Tata Institute of Fundamental Research, Bombay 400 005, India
\item[\ne] Northeastern University, Boston, MA 02115, USA
\item[\bucharest] Institute of Atomic Physics and University of Bucharest,
     R-76900 Bucharest, Romania
\item[\budapest] Central Research Institute for Physics of the 
     Hungarian Academy of Sciences, H-1525 Budapest 114, Hungary$^{\ddag}$
\item[\mit] Massachusetts Institute of Technology, Cambridge, MA 02139, USA
\item[\debrecen] KLTE-ATOMKI, H-4010 Debrecen, Hungary$^\P$
\item[\florence] INFN Sezione di Firenze and University of Florence, 
     I-50125 Florence, Italy
\item[\cern] European Laboratory for Particle Physics, CERN, 
     CH-1211 Geneva 23, Switzerland
\item[\wl] World Laboratory, FBLJA  Project, CH-1211 Geneva 23, Switzerland
\item[\geneva] University of Geneva, CH-1211 Geneva 4, Switzerland
\item[\hefei] Chinese University of Science and Technology, USTC,
      Hefei, Anhui 230 029, China$^{\triangle}$
\item[\seft] SEFT, Research Institute for High Energy Physics, P.O. Box 9,
      SF-00014 Helsinki, Finland
\item[\lausanne] University of Lausanne, CH-1015 Lausanne, Switzerland
\item[\lecce] INFN-Sezione di Lecce and Universit\'a Degli Studi di Lecce,
     I-73100 Lecce, Italy
\item[\lyon] Institut de Physique Nucl\'eaire de Lyon, 
     IN2P3-CNRS,Universit\'e Claude Bernard, 
     F-69622 Villeurbanne, France
\item[\madrid] Centro de Investigaciones Energ{\'e}ticas, 
     Medioambientales y Tecnolog{\'\i}cas, CIEMAT, E-28040 Madrid,
     Spain${\flat}$ 
\item[\milan] INFN-Sezione di Milano, I-20133 Milan, Italy
\item[\moscow] Institute of Theoretical and Experimental Physics, ITEP, 
     Moscow, Russia
\item[\naples] INFN-Sezione di Napoli and University of Naples, 
     I-80125 Naples, Italy
\item[\cyprus] Department of Natural Sciences, University of Cyprus,
     Nicosia, Cyprus
\item[\nymegen] University of Nijmegen and NIKHEF, 
     NL-6525 ED Nijmegen, The Netherlands
\item[\caltech] California Institute of Technology, Pasadena, CA 91125, USA
\item[\perugia] INFN-Sezione di Perugia and Universit\'a Degli 
     Studi di Perugia, I-06100 Perugia, Italy   
\item[\cmu] Carnegie Mellon University, Pittsburgh, PA 15213, USA
\item[\prince] Princeton University, Princeton, NJ 08544, USA
\item[\rome] INFN-Sezione di Roma and University of Rome, ``La Sapienza",
     I-00185 Rome, Italy
\item[\peters] Nuclear Physics Institute, St. Petersburg, Russia
\item[\potenza] INFN-Sezione di Napoli and University of Potenza, 
     I-85100 Potenza, Italy
\item[\salerno] University and INFN, Salerno, I-84100 Salerno, Italy
\item[\ucsd] University of California, San Diego, CA 92093, USA
\item[\santiago] Dept. de Fisica de Particulas Elementales, Univ. de Santiago,
     E-15706 Santiago de Compostela, Spain
\item[\sofia] Bulgarian Academy of Sciences, Central Lab.~of 
     Mechatronics and Instrumentation, BU-1113 Sofia, Bulgaria
\item[\korea]  Laboratory of High Energy Physics, 
     Kyungpook National University, 702-701 Taegu, Republic of Korea
\item[\alabama] University of Alabama, Tuscaloosa, AL 35486, USA
\item[\utrecht] Utrecht University and NIKHEF, NL-3584 CB Utrecht, 
     The Netherlands
\item[\purdue] Purdue University, West Lafayette, IN 47907, USA
\item[\psinst] Paul Scherrer Institut, PSI, CH-5232 Villigen, Switzerland
\item[\zeuthen] DESY, D-15738 Zeuthen, 
     FRG
\item[\eth] Eidgen\"ossische Technische Hochschule, ETH Z\"urich,
     CH-8093 Z\"urich, Switzerland
\item[\hamburg] University of Hamburg, D-22761 Hamburg, FRG
\item[\taiwan] National Central University, Chung-Li, Taiwan, China
\item[\tsinghua] Department of Physics, National Tsing Hua University,
      Taiwan, China
\item[\S]  Supported by the German Bundesministerium 
        f\"ur Bildung, Wissenschaft, Forschung und Technologie
\item[\ddag] Supported by the Hungarian OTKA fund under contract
numbers T019181, F023259 and T024011.
\item[\P] Also supported by the Hungarian OTKA fund under contract
  numbers T22238 and T026178.
\item[$\flat$] Supported also by the Comisi\'on Interministerial de Ciencia y 
        Tecnolog{\'\i}a.
\item[$\sharp$] Also supported by CONICET and Universidad Nacional de La Plata,
        CC 67, 1900 La Plata, Argentina.
\item[$\diamondsuit$] Also supported by Panjab University, Chandigarh-160014, 
        India.
\item[$\triangle$] Supported by the National Natural Science
  Foundation of China.
\end{list}
}
\vfill

%%% Local Variables: 
%%% mode: latex
%%% TeX-master: t
%%% End:

\newpage

%
%%%%%%%%%%%%%%%%%%%%%%%%%%%%%%%%%%%%%%%%%%%%%%%%%%%%%%%%%%%%%%%%%%%%%%%%%%%%%%%
% Bibliography
%%%%%%%%%%%%%%%%%%%%%%%%%%%%%%%%%%%%%%%%%%%%%%%%%%%%%%%%%%%%%%%%%%%%%%%%%%%%%%
%
% Style file to use with mcite.
% Use l3style with just cite.

\bibliographystyle{/l3/paper/biblio/l3stylem}

\clearpage

\begin{table}[t]
\centering
\begin{tabular}{|c|c|c|c|c|}
\hline
\multicolumn{1}{|c|}{reference} & \multicolumn{2}{|c|}{MC, no BEC} & \multicolumn{2}{|c|}{$(+,-)$ pairs} \\
\hline
 channel & fully-hadronic & semi-hadronic         & fully-hadronic & semi-hadronic \\
\hline
\hline
 $\gamma$  & $0.91\pm0.02\pm0.02$ &$\phantom{-}0.94\pm0.01\pm0.02$ & $0.93\pm0.01\pm0.02$ & $0.89\pm0.01\pm0.03$ \\
 $\lambda$ & $0.55\pm0.04\pm0.08$ &$\phantom{-}0.70\pm0.06\pm0.05$ & $0.48\pm0.05\pm0.08$ & $0.64\pm0.07\pm0.06$ \\
 $R$ (fm)  & $0.56\pm0.04\pm0.06$ &$\phantom{-}0.64\pm0.05\pm0.06$ & $0.71\pm0.04\pm0.05$ & $0.75\pm0.05\pm0.06$ \\
 $\delta$  & $0.06\pm0.02\pm0.06$ &$          -0.01\pm0.01\pm0.06$ & $0.07\pm0.01\pm0.05$ & $0.07\pm0.02\pm0.05$ \\
 $\chi^2$/ndf  & 32/31 &41/31 & 29/31 & 35/31 \\
\hline
\end{tabular}
\icaption{Values of the fit parameters
 $\gamma$, $\lambda$, $R$  and $\delta$, eq.\,(\ref{fitg}), for the
 fully hadronic and the semi-hadronic \PW\PW\ events.
 Two different reference samples are used:
 \KORALW\ without BEC, eq.~(\ref{refer0}),
 and unlike-sign particle pairs, eq.~(\ref{refer1}).
 The first error is statistical, the second systematic.
\label{table1}}
\end{table}

\begin{table}
\centering
\begin{tabular}{|c|c|}
\hline
 source & contribution \\
\hline
 background fraction & 0.0021 \\
 other Monte Carlo (\PYTHIA, BES or no BE) & 0.0060 \\
 allowing CR in the reference sample & 0.0024 \\
 fit range & 0.0012 \\
 rebinning (40 $\rightarrow$ 80 \MeV) & 0.0024 \\
 removing $(1+\epsilon Q)$ from the fit & 0.0011 \\
 track selection & 0.0073 \\
 event selection & 0.0046 \\
 mixing & 0.0040 \\
 neural net output cut & 0.0039 \\
 energy calibration & 0.0014 \\
 influence of $\tau$ channel & 0.0076 \\
\hline
 total systematic uncertainty & 0.015\pho \\
\hline
\end{tabular}
\icaption{Contributions to the systematic uncertainty of the $\Lambda$ parameter.
Explanation of the sources are in the text.\label{table2}}
\end{table}

\clearpage

\end{document}